\documentclass[reprint,aps,pra]{revtex4-1}

\usepackage{graphicx}
\usepackage{dcolumn}
\usepackage{bm}
\usepackage{amsmath}
\usepackage{amsthm}
\usepackage{amssymb}
\usepackage{float}
\usepackage{hyperref}
\usepackage{color}
\usepackage{epstopdf}
\usepackage{cleveref}
\usepackage[svgnames]{xcolor}
\hypersetup{hidelinks,colorlinks=true,allcolors=DarkBlue}

\newcommand{\HilbSp}[1]{\mathcal{#1}}
\newcommand{\X}{{\it X}}
\newcommand{\diag}{\mathrm{diag}}
\newcommand{\tom}{\mathrm{tom}}
\newcommand{\opt}{\mathrm{opt}}
\newcommand{\sym}{\mathrm{sym}}
\newcommand{\ket}[1]{|#1\rangle}
\newcommand{\bra}[1]{\langle#1|}
\newcommand{\kb}{k_\mathrm{B}}
\newcommand{\domega}{\Delta\omega}
\newcommand{\ketbra}[2]{\ket{\tilde{#1},\tilde{#2}}\langle\tilde{#1},\tilde{#2}|}

\newcommand{\ct}{\cos{{\theta}/{2}}}
\newcommand{\st}{\sin{{\theta}/{2}}}
\newcommand{\tqb}{\mathrm{2qb}}
\newcommand{\Ptqb}{\hat{\Pi}^\tqb}
\newcommand{\ketbraa}[2]{\ket{#1,#2}\bra{#1,#2}}
\newcommand{\sign}[1]{\mathrm{sign}\left(#1\right)}

\begin{document}

\preprint{APS/123-QED}

\title{Tomographic discord for a system of two coupled nanoelectric circuits}
\author{A.\,K.\,Fedorov$^{1,2,*}$}
\author{E.\,O.\,Kiktenko$^{2,3}$}
\author{O.\,V.\,Man'ko$^{4}$}
\author{V.\,I.\,Man'ko$^{4}$}
\affiliation
{
\mbox{$^{1}$Russian Quantum Center, Skolkovo, Moscow 143025, Russia}
\mbox{$^{2}$Bauman Moscow State Technical University, Moscow 105005, Russia}
\mbox{$^{3}$Geoelectromagnetic Research Center of Schmidt Institute of Physics of the Earth,}
\mbox{Russian Academy of Sciences, Troitsk, Moscow Region 142190, Russia}
\mbox{$^{4}$P.\,N. Lebedev Physical Institute, Russian Academy of Sciences, Moscow 119991, Russia}
}

\date{\today}

\begin{abstract}
We consider quantum correlations and quantum discord phenomena for two-qubit states with $\X$-type density matrices in the tomographic framework of quantum mechanics. 
By introducing different measurements schemes, we establish the relation between tomographic approach to quantum discord, symmetric discord, and measurement-induced disturbance.
In our consideration, $\X$-states appear as approximations of ground and low temperature thermal states of two coupled harmonic oscillators realized by nanoelectric $LC$-circuits.
Possibilities for control the amounts of correlations and entropic asymmetry due to variation of the frequency detuning and the coupling constant are also considered.

\begin{description}
\item[PACS numbers]
03.65.Wj, 03.65.-w, 03.67.-a
\end{description}
\end{abstract}
                              
\maketitle

\section{Introduction}

Inspiring experimental progress on creation, manipulation, and characterization of individual quantum objects has explored new frontiers in quantum science and technologies \cite{Lukin}. 
Due to the intriguing properties of quantum systems, they can be viewed as a potential platform for 
ultra-sensitive metrology \cite{Ye}, 
unconditionally secure communications \cite{Gisin}, 
high-efficient information processing \cite{Ladd}.
and simulation of complex quantum systems \cite{Cirac}.

However, implementation of quantum devices is challenging, because even bipartite quantum systems exhibit a non-trivial behavior and correlation properties.
A shining example is that of quantum discord, demonstrating a specific type of quantum correlations in bipartite quantum systems 
\cite{discord, discord2, discord3, discord4, discord5,discord6, discord7, discord8, discord9, discord10, discord11, discord12, discord13, discord14,SymDisc,Neumann}.
Quantum discord is a measure of quantum correlations based on subtraction of locally accessible information from the total amount of the quantum mutual information. 
The exploration of quantum discord has inspired a new wave of research of nonlocal properties of separable quantum states \cite{separable} 
and their applications for realizing of quantum algorithms \cite{separable2, separable3}.

The concept of quantum discord is associated to quantum measurements. 
A powerful technique for complete experimental characterization of quantum states and processes in terms of non-negative probability distribution function is quantum tomography \cite{Lvovsky}. 
The tomographic approach was introduced for systems with continuous \cite{Manko1} and discrete variables \cite{Manko2}. 
Clearly, the tomographic approach generalizes the Shannon information theory on the quantum domain in a very natural way.
In the past few decades, a wide class of problems in quantum information theory, {\it e.g.}, revealing new inequalities for Shannon \cite{Shannon} and R\'enyi entropies \cite{MAManko1, FKMM, MAManko2}, 
tomographic approach to the Bell-type inequalities \cite{Bell}, quantumness tests \cite{Filippov1}, quantum correlations and quantum discord \cite{MAManko2, Manko8, Kiktenko1,OVManko1}, 
has been investigated in detail \cite{Review}.  

\begin{figure} \label{fig:1}
\includegraphics[width=0.575\linewidth]{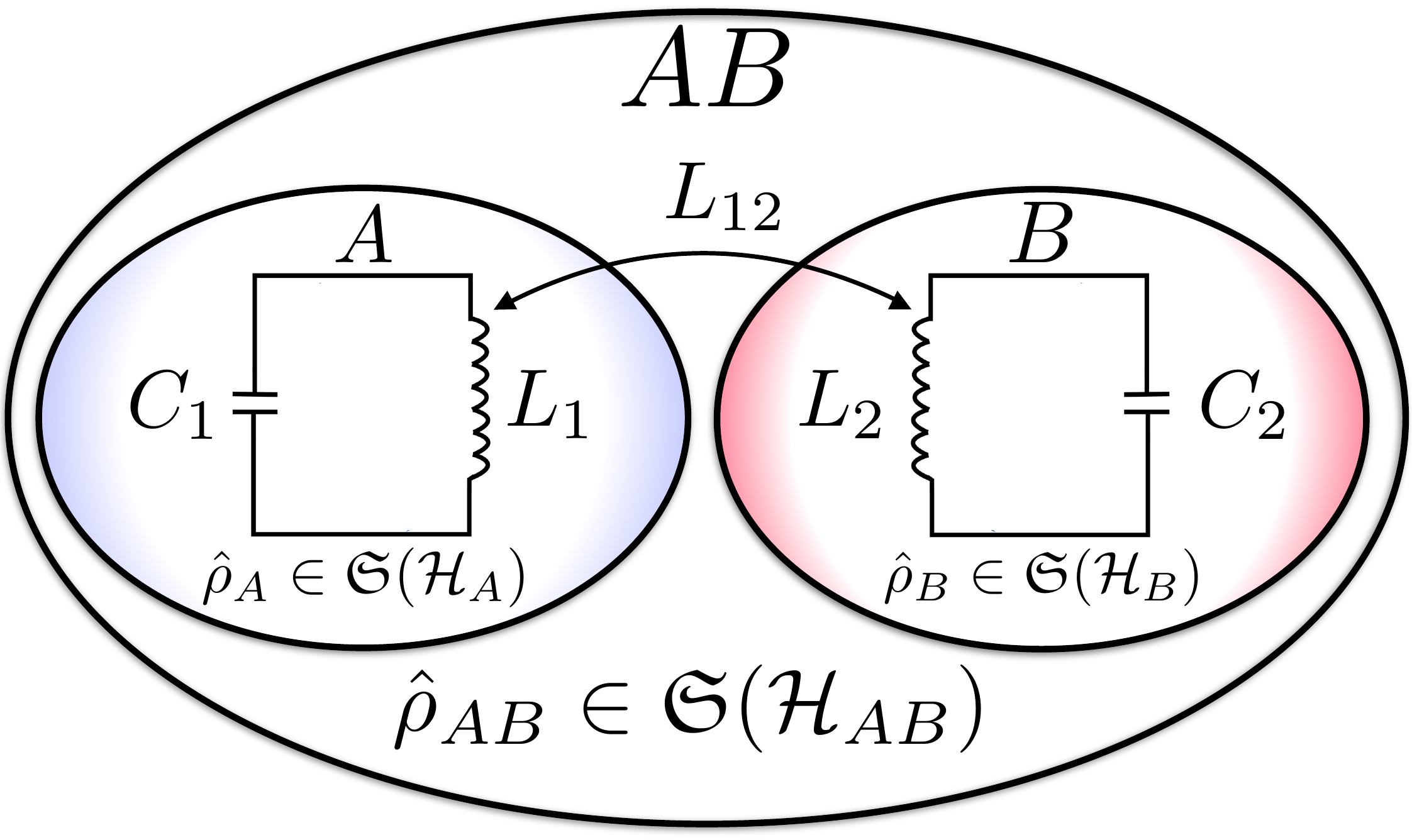}
\vskip -4mm
\caption
{
Model a two-qubit system with $X$-type of the density matrix $\rho_{AB}$. 
These states appear as approximations of ground and low temperature thermal states of two nanoelectric $LC$-circuits coupled through mutual inductance $L_{12}$.
}
\end{figure}

In the quantum optics domain, the field quadratures of an electromagnetic mode play a role of canonical positions and momentums operators \cite{Leonhardt}.
Using optical homodyne detection \cite{Lvovsky,Beck,Beck2}, the uncertainty relations, the entropic inequalities and quantum discord have been experimentally verified \cite{Porzio, Bellini,RahimiKeshari}. 
It should be noted, however, that tomographic approach is applicable to arbitrary quantum systems whose Hilbert space is isomorphic the Hilbert space of the harmonic oscillator. 

Recently, the possibility of applying quantum tomography for description of quantum states of current and voltage in a quantum nanoelectric circuits with the Josephson junction has been demonstrated \cite{OVManko2,OVManko3}. 
Using symplectic tomography \cite{Manko1}, examples of the Gaussian states of circuits with the Josephson junction and two coupled resonant circuits of high quality have been considered~\cite{OVManko2}. 
Tomographic expressions for the Shannon entropy, mutual information, fidelity, and purity of the quantum states of a nanoelectric circuit have been obtained \cite{OVManko3}.

At the same time, substantial advances in the manufacturing of quantum nanoelectric circuits and superconducting quantum interference devices have been achieved \cite{Maklin}. 
Due to there being sufficiently low dissipation in nanoelectric circuits, they are promising candidates for providing scalable interfaces between classical circuits and the quantum counterparts \cite{sca}.
The realization of simple two-qubit algorithms using a superconducting quantum processor has been demonstrated \cite{scc}.

In addition to their importance in applications \cite{Maklin}, nanoelectric circuits and Josephson junctions with time-varying parameters provide useful setups for a model of a parametric quantum oscillator
\cite{DodonovManko1, OVManko4, Manko4, Dodonov1, Dodonov2, Dodonov3, Dodonov3}. 
The significance of this model has been shown in theoretical \cite{Dodonov1, Dodonov2, Dodonov3,DodonovLozovik,Zeilinger} and experimental \cite{Wilson} studies of the Casimir effect \cite{Casimir}. 

In the present work, we consider quantum correlations, quantum discord, and entropic asymmetry for a class of states with $\X$-type density matrices using quantum tomography.
We are interested in several important issues.
The considered class of $\X$-states density matrices is of particular interest because there exists an analytical formula for quantum discord \cite{discord5,discord6,discord7,discord8}, 
which is either precise for a huge subclass of $X$-states or gives a sufficiently small error \cite{discord9}.
A first important problem is that of revealing the connection between the tomographic approach to quantum discord \cite{MAManko2, Manko8,Kiktenko1} and discord related measures~\cite{discord}.
Using various kinds of measurement schemes, we establish the relations among the tomographic discord, symmetric discord based on the von
Neumann measurements, and measurement-induced disturbance, we establish the relation between tomographic discord, symmetric discord based on the von Neumann measurements, and measurement-induced disturbance \cite{discord4}.
Also, we propose an analytical formula for quantum discord as well as examining it for the set of randomly generated two-qubit states. 
Furthermore, we combine our consideration with quantum causal analysis \cite{Kikt, Kikt2, Kikt3, Kikt4}, 
which allows us to reveal significant properties in the entropic asymmetry of states with $\X$-type density matrices \cite{Zyczkowski}.

The connection of quantum discord with well controllable and easy implementable physical systems opens a way for its experimental investigations.
There are several notable proposals to realize quantum states with $\X$-type density matrix for biparticle systems in experiments: 
spin-$1/2$ particles with the XY Hamiltonian in the external magnetic field \cite{discord10};
two-level atoms, driven by a laser field interacting \cite{discord11};
coupled superconducting circuits, based on the Josephson junctions \cite{discord12} and others.
Recently, it has been demonstrated that two-qubit $X$-states emerge in the ground states of a large class of Hamiltonians, 
including XY model, XXZ model, and transverse field Ising model \cite{discord13}. 
In view of the aforementioned progress in the manufacturing, characterization, and manipulation of superconducting circuits \cite{Maklin,sca,scc,Menzel,Mallet},  
we suggest obtaining $X$-states as approximations of ground and low temperature thermal states for a system of two coupled harmonic oscillators, realized using nanoelectric $LC$-circuits (Fig. 1).
Thus, the second question concerns the controllability of the amounts of quantum correlations and entropic asymmetry through variation of the frequency detuning and the coupling constant of two coupled nanoelectric $LC$-circuits. 
We are also interested in the robustness of correlation properties with respect to the controllable parameters of the physical system and environment.

The paper is organized as follows.
We review quantum tomography for discrete variables, tomographic information measures, tomographic quantum discord, and causal analysis in Section \ref{Tomography}. 
Using various measurement schemes, we establish the connection between tomographic discord and discord related measures.
In Section \ref{Schemes}, we study various measurement schemes for $\X$-states. 
Consequently, we suggest an analytical formula as well as verify it on the set of randomly generated two-qubit states. 
$\X$-states are obtained as approximations of ground and low temperature thermal states for two coupled nanoelectric $LC$-circuits in Section \ref{Circuits}.  
We investigate correlation properties of the two-qubit system with $\X$-type density matrix as function of controllable parameters of our physical system: the frequency detuning and the coupling constant of circuits. 
Finally, we give conclusions and prospectives in Section \ref{Conclusion}.

\section{Quantum tomography}\label{Tomography}

In general, quantum states are described via the density operator $\hat\rho\in\mathfrak{S}(\mathcal{H})$, 
where $\mathfrak{S}(\mathcal{H})$ is the set of positive operators of unit trace ${\rm Tr}{\hat{\rho}}=1$ in a Hilbert space $\mathcal{H}$.

For the finite-dimensional Hilbert space case, we can introduce quantum tomograms as follows:
\begin{eqnarray}\label{eq:defenitiond}
	\mathcal{T}(U)=\left\{ \mathcal{T}_m(U) \right\}=
	\left\{\langle{m}|U\hat\rho{U}^{\dagger}|{m}\rangle\right\},
\end{eqnarray}
where $\{\ket{m}\}$ is the complete set of an orthonormal vectors, representing a measurement basis, and $U\in{\rm SU}(N)$ is the unitary matrix.

Having a fair probability distribution function, quantum tomograms are positive and normalized:
$$
	{\mathcal{T}_m(U)}\geq0, \qquad \sum\nolimits_m{\mathcal{T}_m(U)}=1.
$$

For qubit systems with $\dim\mathcal{H}=2$ and $m=\ket{0},\ket{1}$, generic form of definition (\ref{eq:defenitiond}) reduces to ${\rm SU}(2)$ case \cite{Manko2}.
Thus, matrix $U\in{\rm SU}(2)$ can be parametrized in the form
\begin{equation} \label{eq:RotOp}
	U{=}U\left( \theta,\phi \right){=}
	{\begin{pmatrix}
		 \ct & \st \\
	 	-\st & \ct 
	\end{pmatrix}}
	{\begin{pmatrix}
		e^{i\phi/2} & 0 \\
		0 & e^{-i\phi/2}
	\end{pmatrix}},
\end{equation}
where $\theta$ and $\phi$ corresponds to the Bloch sphere rotation.

\subsection{Mutual information}\label{sec:4}

Let us consider a bipartite state $AB$ with the following density operator:
$$
	\hat{\rho}_{AB}\in\mathfrak{S}(\mathcal{H}_{AB}=\mathcal{H}_{A}\otimes\mathcal{H}_{B}),
$$
where the density operators of subsystems have the forms
$$
	\hat{\rho}_{A}=\mathrm{Tr}_{B}\hat{\rho}_{AB}\in\mathfrak{S}(\mathcal{H}_{A}), \qquad \hat{\rho}_{B}=\mathrm{Tr}_{A}\hat{\rho}_{AB}\in\mathfrak{S}(\mathcal{H}_{B}).
$$

In quantum information theory, the full amount of correlations between subsystems is measured by the quantum mutual information:
\begin{equation}\label{eq:QMutInf}
	I=S_A+S_B-S_{AB}, 
\end{equation}
where $S_A$, $S_B$ and $S_{AB}$ are the von Neumann entropies, given by the general expression:
\begin{equation}\label{eq:QEnropy}
	S=-\mathrm{Tr}\left(\hat{\rho}\log\hat{\rho}\right)
\end{equation}
Here, we take the logarithm to base 2.

In the tomographic picture of quantum mechanics, the bipartite quantum state can be described by tomogram in the following form:
$$
	\mathcal{T}_{AB}(U_A\otimes U_B)=\left\{\mathcal{T}_{AB_{ij}}(U_A\otimes U_B)\right\},
$$
where $U_A$ and $U_B$ are operators, which describe local orthogonal projective measurements of subsystems $A$ and $B$. 
Then, tomograms of subsystems take the form
\begin{eqnarray}
	\mathcal{T}_{A}(U_A)=\{\mathcal{T}_{A_i}(U_A)\}
	=\left\{\sum\nolimits_j\mathcal{T}_{AB_{ij}}(U_A\otimes U_B)\right\}, \nonumber \\ 
	\mathcal{T}_{B}(U_B)=\{\mathcal{T}_{B_j}(U_B)\}
	=\left\{\sum\nolimits_i\mathcal{T}_{AB_{ij}}(U_A\otimes U_B)\right\}. \nonumber
\end{eqnarray}
Being, on the one hand, probability distribution functions, and, on the other hand, applicable for the description of quantum states, quantum tomograms allow us to study both classical and quantum correlations between subsystems in bipartite quantum states.

In this way, we can introduce the tomographic Shannon entropy \cite{Shannon} calculated for the tomogram $\mathcal{T}(U)$:
\begin{equation} \label{eq:ShEntr}
	H(U)=-\sum\nolimits_m {\mathcal{T}_{{m}}(U) \log{\mathcal{T}_{{m}}(U)}}.
\end{equation}
	
Thus, one can describe the amount of observable by local measurements correlations in the bipartite system $AB$ via the tomographic mutual information,
\begin{equation}\label{eq:CMutInf}
	J(U_A,U_B)=H_A(U_A)+H_B(U_B)-H_{AB}(U_A\otimes U_B),
\end{equation}
where $H_A(U_A)$, $H_B(U_B)$ and $H_{AB}(U_A\otimes U_B)$ are tomographic Shannon entropies calculated using 
Eq. \eqref{eq:ShEntr} for $\mathcal{T}_A(U_A)$, $\mathcal{T}_B(U_B)$ and $\mathcal{T}_{AB}(U_A\otimes U_B)$, correspondingly.

Expression (\ref{eq:CMutInf}) shows that the tomographic mutual information is a straightforward analog of (\ref{eq:QMutInf}),
while tomographic Shannon entropy (\ref{eq:ShEntr}) generalizes von Neumann entropy (\ref{eq:QEnropy}).
The value for tomographic mutual information (\ref{eq:CMutInf}), clearly, depends on operators $U_A$ and $U_B$. 

\subsection{The discord}

The conventional approach \cite{discord} is to define quantum discord as the difference between the total correlations (\ref{eq:QMutInf}) 
and the classical correlations obtained after a measurement performed on one subsystem ({\it e.g.}, on the subsystem $B$):
\begin{equation}\label{eq:Disc}
	D^{(B)}=I-\max_{\{\Pi_b\}}J^{(B)}_{\{\Pi_b\}}\geq0,
\end{equation}
with $J^{(B)}_{\{\Pi_b\}}$ being the quantum mutual information calculated by Eq. (\ref{eq:QMutInf}) for the state
$$
	\rho_{AB}^{(B)}=\sum_b M_b \rho_{AB} M_b^\dagger, \qquad \Pi_b=M_b^\dagger M_b
$$
where the quantity $\Pi_b$ introduced is the positive-operator valued measure (POVM) in the Hilbert space $\HilbSp{H}_B$. 

On the other hand, from definitions (\ref{eq:QMutInf}) and (\ref{eq:CMutInf}), one can construct the tomographic discord as a tomographic measure of quantumness of bipartite state correlations,
\begin{equation}\label{eq:TomDisc}
	D(U_A,U_B)=I-J(U_A, U_B).
\end{equation}
Tomographic measure (\ref{eq:TomDisc}) resembles the concept of symmetric quantum discord \cite{SymDisc}. 
The difference is as follows.
First, in definition (\ref{eq:Disc}) of quantum discord, a measurement is performed only on the one subsystem.
Second, the measurement in (\ref{eq:TomDisc}) is described by set of orthogonal projectors (von Neumann measurements).
Finally, there is no maximization procedure in definition (\ref{eq:TomDisc}).

Due to the dependence of tomographic mutual information (\ref{eq:CMutInf}) and, therefore, expression (\ref{eq:TomDisc}), on unitary operators $U_A$ and $U_B$,
the amount of observed correlations (or quantumness) directly depends on the measurements.

We consider three different approaches to the unambiguous definition of that operators.

\subsubsection{The optimal measurement scheme}

The first approach follows directly from the definition of symmetric quantum discord, related to the von Neumann measurements \cite{Neumann}. 
The idea is to consider measurement operators $U_A$ and $U_B$ in such way that they maximize tomographic mutual information (\ref{eq:CMutInf}),
\begin{equation} \label{eq:OptS}
	\left(U_A^\opt,U_B^\opt\right)=\arg\max_{U_A,U_B}J(U_A, U_B).
\end{equation}
The resulting tomogram of the state,
\begin{equation}\label{eq:OptTom}
	\mathcal{T}_{AB}^\opt=\mathcal{T}_{AB}\left(U_A^\opt\otimes U_B^\opt\right)
\end{equation}
makes quantity (\ref{eq:TomDisc}) to attain the minimal possible value:
\begin{equation} \label{eq:OptTomDisc}
	D^\opt=I-J^\opt, \quad J^\opt=J\left(U_A^\opt, U_B^\opt\right)
\end{equation}
Due to existence of the maximization procedure in Eq. (\ref{eq:OptS}), we label this approach to choice of unitary operators $U_A$ and $U_B$ with corresponding correlation measure as optimal tomographic discord.

\subsubsection{The diagonalizing measurement scheme}

An alternative natural approach is to consider unitary operators $U_A$ and $U_B$ in such a way that the density matrices of the subsystems keeps to be undisturbed after measurement \cite{Manko8}. 
In other words, the density matrices of the subsystems after action of the operators become diagonal in the measurement basis.

Applying unitary transformations with these operators for the density matrices of the subsystems, we find that the Shannon entropies of subsystems become equal to the von Neumann entropies,
$$
	H_A\left(\widetilde{U}_A^\diag\right)=S_A, \qquad H_B\left(\widetilde{U}_B^\diag\right)=S_B.
$$
We use tilde and superscript ``diag'' for notation of these operators.

However, there is still ambiguity as regards their choice.
For example, for maximally entangled states, density matrix of the subsystems keeps to be diagonal (and proportional to identity matrix) under any possible rotation operator.

Therefore, we consider the set of operators given by:
\begin{equation} \label{eq:DiagS}
	\left(U_A^\diag,U_B^\diag\right)=\arg\max_{\widetilde{U}_A^\diag,\widetilde{U}_B^\diag}J\left(\widetilde{U}_A^\diag,\widetilde{U}_B^\diag\right),
\end{equation}
not only to make density matrices of the subsystems to be diagonal, but also to maximize the level of correlations described by Eq. (\ref{eq:CMutInf}).

We denote the corresponding tomogram as follows:
\begin{equation}\label{eq:DiagTom}
	\mathcal{T}_{AB}^\diag=\mathcal{T}_{AB}\left(U_A^\diag\otimes U_B^\diag\right).
\end{equation}
As a result, one can introduce diagonalizing tomographic discord in the following form:
\begin{multline} \label{eq:DiagTomDisc}
	D^\diag=I-J^\diag=H_{AB}(U_A^\diag\otimes U_B^\diag)-S_{AB}, \\
	J^\diag=J(U_A^\diag,U_B^\diag).
\end{multline}	
Being introduced in tomographic framework in Ref.~\cite{Manko8}, this quantity is also well-known as measurement-induced disturbance \cite{discord4}. 
It should be noted that (\ref{eq:DiagTomDisc}) was named  ``tomographic discord'' in work \cite{Kiktenko1}, 
however, in this work we prefer to use the term ``diagonalizing tomographic discord'' because we also consider another approaches, which are inherently tomographic as well.

\subsubsection{The symmetrizing measurement scheme}

Here, we point out one more auxiliary approach. 
This approach arises from the fact that for certain class of states (notably, for $\X$-states), 
it is useful to consider the unitary operators such that tomograms of the subsystems become uniform distributions, {\it e.g.},
for two-qubit states, corresponding tomograms read:
$$
	\mathcal{T}_A\left(\widetilde{U}_A^\sym\right)=\mathcal{T}_B\left(\widetilde{U}_B^\sym\right)=\{1/2,1/2\}.
$$
Consequently, introduced operators transform Shannon entropies [see Eq. (\ref{eq:ShEntr})] of the subsystems to be equal to their maximum possible values:
$$
	H_A\left(\widetilde{U}_A^\sym\right)=H_B\left(\widetilde{U}_B^\sym\right)=1.
$$

Like in giving the definitions (\ref{eq:OptTom}) and (\ref{eq:DiagS}), we add a requirement that tomographic mutual information (\ref{eq:CMutInf}) attains its maximum possible value:
\begin{equation} \label{eq:SymS}
	\left(U_A^\sym,U_B^\sym\right)=\arg\max_{\widetilde{U}_A^\sym, \widetilde{U}_B^\sym}J\left(\widetilde{U}_A^\sym,\widetilde{U}_B^\sym\right),
\end{equation}
Finally, the corresponding measure reads:
\begin{multline} \label{eq:SymTomDisc}
	D^\sym=I-J^\sym=I+H_{AB}(U_A^\sym\otimes U_B^\sym)-2, \\
	J^\sym=J(U_A^\sym\otimes U_B^\sym).
\end{multline}
We label this approach as symmetrizing discord.

\subsection{Entropic asymmetry}

Entropic asymmetry of particular mixed states is an another interesting and important issue \cite{Zyczkowski}. 
Due to such asymmetry, decoherence acting on different parties leads to different rates of correlations decay.
In other words, the question about robustness of parties appears.

For that purpose, quantum causal analysis~\cite{Kikt,Kikt2,Kikt3,Kikt4} and its tomographic generalization \cite{Kiktenko1} have been proposed.  
This approach was successfully implemented to two-\cite{Kikt2} and three-\cite{Kikt3} qubit states and atom-field interactions \cite{Kikt4}.
We note that pure bipartite states always have equal von Neumann entropies of the subsystems due to Schmidt decomposition.
In this way, they are not of interest. 

Quantum causal analysis is based on a pair of the independence functions \cite{Kikt}:
$$
	i_{A|B}=\frac{S_{A}-I}{S_A}, \qquad i_{B|A}=\frac{S_{B}-I}{S_B},
$$
which can be used for the measure of entropic asymmetry,
\begin{equation}\label{eq:AsMes}
	d_{AB}=i_{A|B}-i_{B|A}=I\frac{S_A-S_B}{S_AS_B}.
\end{equation}
This measure of asymmetry (\ref{eq:AsMes}) has the following useful properties:
(i) positive values of $d_{AB}$ correspond to a case where the first subsystem plays a decisive role in correlation compared to the second one 
(in the causal analysis the first subsystem get label the ``cause'', while the second one is called the ``effect'');
(ii) $d_{AB}=-d_{BA}$, {\it i.e.}, the sign of $d_{AB}$ defines the direction of the asymmetry;
(iii) the magnitude $|d_{AB}|$ corresponds to the extent of asymmetry between roles of subsystems in correlations.

In turn, tomographic approach to the amount of asymmetry \cite{Kiktenko1} is relied on the tomographic independence functions,
\begin{equation}\label{eq:TomInFun}
	\begin{aligned}
		i_{A|B}^\tom(U_A,U_B)=\frac{H_{A}(U_A)-J(U_A,U_B)}{H_A(U_A)}, \\ i_{B|A}^\tom(U_A,U_B)=\frac{H_{B}(U_B)-J(U_A,U_B)}{H_B(U_B)},
	\end{aligned}
\end{equation}
with corresponding measure of asymmetry in the form:
\begin{equation}\label{eq:TomAsMes}
	d_{AB}^\tom(U_A,U_B)=i_{A|B}^\tom(U_A,U_B)-i_{B|A}^\tom(U_A,U_B).
\end{equation}

Substituting in (\ref{eq:TomInFun}) the unitary operators $U_A$ and $U_B$ in the form of (\ref{eq:OptS}), (\ref{eq:DiagS}), and (\ref{eq:SymS}), 
we obtain three definite values of tomographic asymmetry: 
$d_{AB}^\opt$, $d_{AB}^\diag$ and $d_{AB}^\sym$, correspondingly. 
However, the third value is of no interest since we always have the identity $d_{AB}^\sym=0$.

Additionally, we note that diagonalizing scheme has the following important property:
$$
	\frac{d_{AB}^\diag}{d_{AB}}=\frac{J^\diag}{I}=1-\frac{D^\diag}{I}>0.
$$
Thus, this scheme does not change the direction of original asymmetry, but it can decrease its extent.

\section{The tomographic quantum discord for $\X$-states}\label{Schemes}

\begin{figure*}[t]
\begin{centering}
\includegraphics[width=0.95\textwidth]{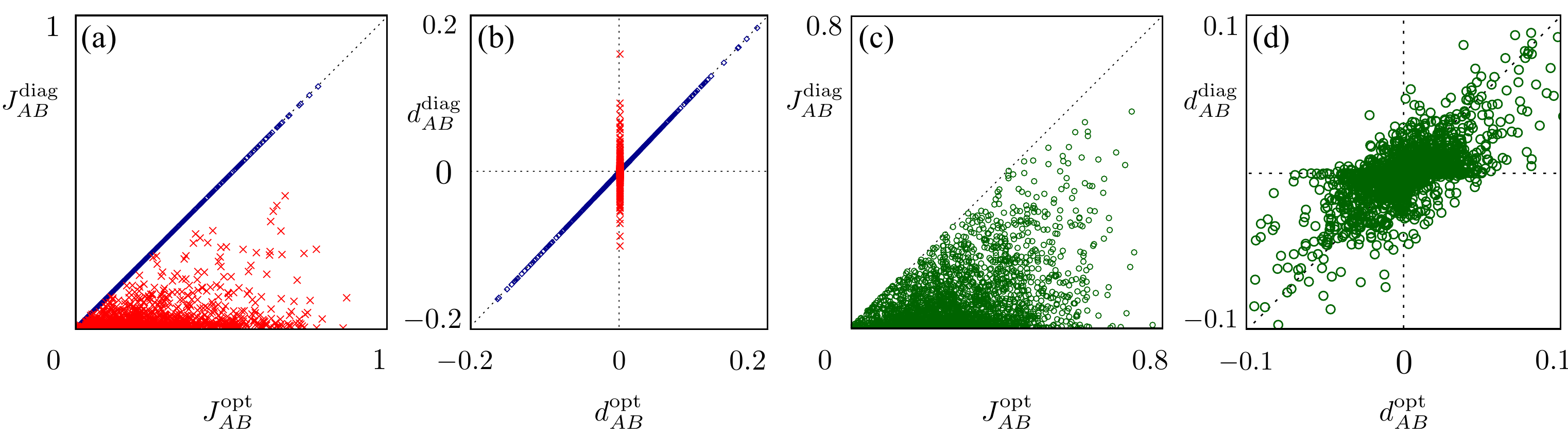}
\end{centering}
\vskip -4mm
\caption
{
The comparison of tomographic mutual information (\ref{eq:CMutInf}) and entropic asymmetry for $3{\times}10^3$ randomly generated $\X$-states [see Appendix IIA]
and $3{\times}10^3$ arbitrary mixed two-qubit states [see Appendix IIB].
In (a) comparison of tomographic mutual information (\ref{eq:CMutInf}) for $\X$-states: 
obtained from diagonalizing (\ref{eq:DiagTomDisc}) measurement scheme {\it vs}. obtained in optimal (\ref{eq:OptTomDisc}) measurement scheme.
In (b) asymmetry measure obtained diagonalizing (\ref{eq:DiagTomDisc}) measurement scheme {\it vs}. the asymmetry obtained in optimal (\ref{eq:OptTomDisc}) measurement scheme.
One can see the separation on tomographically asymmetric states (crosses) and tomographically symmetric states (rhombus).
In (c)-(d) the same quantities for randomly generated states (circles).
} 
\label{fig:RandGen}
\end{figure*}

In the current paper, we give our main attention to the class of two-qubit states with $\X$-type density matrix:
\begin{equation} \label{eq:XState}
	\rho_{AB}^\X=\begin{pmatrix}
	 \rho_{11} && 0 && 0 && \rho_{14} \\
	 0 && \rho_{22} && \rho_{23} && 0 \\
	 0 && \rho_{23} && \rho_{33} && 0 \\
	 \rho_{14} && 0 && 0 && \rho_{44} \\
	\end{pmatrix}.
\end{equation}
The constraints on the elements $\{\rho_{ij}\}$ are the following: 
(i) all diagonal elements are nonnegative and form a unit trace;
(ii) the off-diagonal elements are also nonnegative and their magnitudes are bounded by inequalities:
$$
	\rho_{23}^2\leq \rho_{22}\rho_{33}, \qquad \rho_{14}^2\leq \rho_{11}\rho_{44}.
$$
We note that any hermitian matrix in form (\ref{eq:XState}) with complex off-diagonal elements always can be transformed in the $\X$-type matrix with all element being nonnegative by a suitable choice of the basis \cite{discord6}. 

The $\X$-states have a paramount importance for the concept of quantum discord \cite{discord}.
Recently, for this class of states attract a great deal of interest in the context of a search for the analytical formula for its computation \cite{discord5, discord6, discord7,discord8,discord9}. 

The intention of our work is to study how optimal (\ref{eq:OptTomDisc}), diagonalizing (\ref{eq:DiagTomDisc}), and symmetrizing (\ref{eq:SymTomDisc}) discords relates to each other for the case of $\X$-states (\ref{eq:XState}).

\subsection{Diagonalizing and symmetrizing tomograms}

For $\X$-states, tomogram (\ref{eq:defenitiond}) parametrized by the Bloch sphere rotation angles:
$$
	\mathcal{T}_{AB}=\mathcal{T}_{AB}(\phi_A,\theta_A,\phi_B,\theta_B)
$$ 
can be easily obtained from diagonal elements of the initial density matrix (\ref{eq:XState}) after implementation of the rotation operator 
$U(\phi_A,\theta_A)\otimes U(\phi_B,\theta_B)$ (where both operators have the form (\ref{eq:RotOp}))
The explicit results of our calculations are presented in Appendix I. 

Here, we consider the main entropic properties:

(i) Due to the symmetry considerations, 
we restrict ourself in studying parameters in the following regions:  
$$
	\theta_A,\theta_B \in \left[0,\pi/2\right], \qquad 
	\phi_A,\phi_B \in \left[0,\pi\right].
$$

(ii) The tomograms of the subsystems appear to be functions only of rotation angles $\theta_A$ and $\theta_B$:
$$
	\mathcal{T}_A=\mathcal{T}_A(\theta_A), \qquad
	\mathcal{T}_B=\mathcal{T}_B(\theta_B).
$$ 
Moreover, at $\theta_A{=}\theta_B{=}0$, tomographic Shannon entropies (\ref{eq:ShEntr}) are equal to von Neumann ones (\ref{eq:QEnropy}),
$$
	H_A\left(\theta_A=0\right)=S_A, \qquad 
	H_B\left(\theta_B=0\right)=S_B;
$$
whereas for the condition $\theta_A{=}\theta_B{=}\pi/2$, they attain their maximum values
$$
	H_A\left(\theta_A=\pi/2\right)=H_B\left(\theta_B=\pi/2\right)=1.
$$

(iii) Since the all matrix elements in (\ref{eq:XState}) are real, 
the maximum value of tomographic mutual information $J(\phi_A,\theta_A,\phi_B,\theta_B)$ at fixed values of the angles $\theta_A$ and $\theta_B$ is obtained at $\phi_A=\phi_B=0$.

From (ii) and (iii), it follows directly that the condition $\theta_A{=}\theta_B{=}\phi_A{=}\phi_B{=}0$ corresponds a diagonalizing tomogram with the following form:
\begin{equation} \label{DiagTom}
	\mathcal{T}_{AB}^\diag=\{\rho_{ii}, i=1\ldots4\}.
\end{equation}

Meanwhile, a symmetrizing tomogram is obtained for $\theta_A=\theta_B=\pi/2$ and $\phi_A=\phi_B=0$, and it has the form:
\begin{equation} \label{eq:SymTom}
	\mathcal{T}_{AB}^\sym=\left\{\frac{1}{4}+\kappa,\frac{1}{4}-\kappa,\frac{1}{4}-\kappa,\frac{1}{4}+\kappa\right\}, 
\end{equation}
with the parameter $\kappa=(\rho_{14}+\rho_{23})/2$.
Explicit formulas (\ref{DiagTom}) and (\ref{eq:SymTom}) makes it straightforward to compute diagonalizing (\ref{eq:DiagTomDisc}) and symmetrizing (\ref{eq:SymTomDisc}) discords as well as 
the corresponding asymmetry measure.

\subsection{The optimal tomogram}

To resolve a problem concerning the optimal measurement scheme, we implement a numerical procedure on a set of $3\times10^3$ randomly generated $\X$-states, similar to that in work \cite{Kiktenko1}.
The methodology used for the generation of random $\X$-states is described in Appendix IIA. 

The results concerning the comparison between $D^\diag$ and $D^\opt$, as well as $d_{AB}^\diag$ and $d_{AB}^\opt$, are presented in Fig. 2a and Fig. 2b. 
Analyzing these numerical results, one can conclude that class of all generated $\X$-states separates into to subclasses:

(i) the first subclass with [circles on Fig. 2(a)],
$$
	D^\opt=D^\diag, \qquad d^\opt=d^\diag; 
$$

(ii) the second one with [crosses on Fig. 2(a)],
$$
	D^\opt<D^\diag, \qquad d^\opt=0, \qquad \mathcal{T}_{AB}^\opt=\mathcal{T}_{AB}^\sym.
$$
We conclude that for $\X$-states, the optimal tomogram is either the diagonalizing or the symmetrizing one:
\begin{equation} \label{eq:OptD}
	D^\opt=\min{\left( D^\diag, D^\sym\right)}.
\end{equation}
Thus, one can label the first subclass (i) of $\X$-states as ``tomographically asymmetric'', and the second one (ii) as ``tomographically symmetric'' subclass (see Fig 2b).

We note that this result is quite in the spirit of the analytical formula for the canonical quantum discord for $\X$-states, obtained in Ref. \cite{discord5}, 
where the optimal measurement of {\it one} subsystem only should be performed either along $z$ or $x$ axis of the Bloch sphere ($\sigma_x$ or $\sigma_z$ measurements).
As it has been demonstrated in Ref.~\cite{discord7}, this approach is appropriate not for all variety of $\X$-states.
In our case, the measurement is performed over the {\it both} subsystems. 
Then, the established Eq. (\ref{eq:OptD}) seems to be correct for the whole class of states with $\X$-type density matrix.
As far as we know, for two-qubit case the question about whether orthogonal projective measurements, as compared to POVMs of rank 1, maximize the classical correlations is still open.

Besides, pure biparticle quantum states that are not maximally entangled belong to a tomographically asymmetric subclass of $\X$-states with optimal measurement basis being defined by their Schmidt decomposition 
We note that for maximally entangled states, the diagonalizing and symmetrizing schemes coincide.

Moreover, for pure states the following equalities holds:
\begin{equation}\label{pure}
	D^\opt=D^\diag=D^{(A)}=D^{(B)}=J^\diag=\frac{1}{2}I=\mathcal{E},
\end{equation}
where $D^{(A)}$ and $D^{(B)}$ are canonical discords (\ref{eq:Disc}), obtained from measurements on $A$ and $B$, correspondingly; and $\mathcal{E}$ is the entanglement of formation being just equal to entropies $S_A=S_B$ in this case.

In Fig. 2c and Fig. 2d, we demonstrate results obtained for randomly generated arbitrary mixed two-qubit states.
Generation of random arbitrary mixed two-qubit states is discussed in Appendix IIB. 

One can conclude that in the general case the optimal measurement scheme is different both from diagonal and symmetrizing. 
Thus, we obtain:
$$
	D^\opt \leq \min{\left( D^\diag,D^\sym \right)}.
$$
Moreover, the direction of asymmetry defined by,
$$
	\sign{d_{AB}^\diag}=\sign{d_{AB}}
$$
may be opposite to $\sign{d_{AB}^\opt}$.

Thus, this separation on the subclasses with respect to introduced correlation measure and measures of entropic asymmetry is not universal for all two-qubit states [see Fig. 2d], 
{\it i.e.}, it is a specific feature of $\X$-states.

\section{Ground and thermal states of quantum circuits}\label{Circuits}

In this section, we propose physical realization for states with $\X$-type density matrices as approximations of ground and low temperature thermal states of coupled quantum nanoelectric circuits [see Fig. 1].

The Hamiltonian of a system of two circuits with inductances $L_1$, $L_2$, capacitors $C_1$, $C_2$, coupled via mutual inductance $L_{12}$, reads:
\begin{equation}\label{eq:Hamilt0}
	\hat{\mathcal{H}}=\hat{\mathcal{H}}_{1}+\hat{\mathcal{H}}_{2}+\hat{V}, \,\,  \hat{V}=L_{12}\hat{I}_{1}\hat{I}_{2}, \,\, \left[\hat{I}_i, \hat{Q}_i\right]={\rm i}\hbar/L_i,
\end{equation}
where $\hat{I}$ is the current operator and 
$\hat{Q}$ is the operator of charge on the plates of a capacitor with the standard commutation relation.
Here, the Hamiltonian,
\begin{equation}\label{hamilton0}
	 \quad \hat{\mathcal{H}_j}=\frac{L_j\hat{I}_j^2}{2}+\frac{\hat{Q}_j^2}{2C_j}, \qquad j=1,2,
\end{equation} 
corresponds to a single resonant circuit.

We assume that the energy of thermal fluctuations in circuits is sufficiently smaller than the energy of quanta, 
{\it i.e.}, $\hbar\omega_j>\kb T$, where $T$ is the temperature, $\omega_j=(L_jC_j)^{-1/2}$ is the resonant frequency, and $\kb$ is the Boltzmann constant.
Thus, we can consider circuits as quantum ones. 

Due to the duality between mechanical oscillators and circuits, it is convenient to introduce canonical positions and momenta operators,
\begin{equation} \label{eq:CanonicalV}
	\hat{x}_j=-L_jC_j^{1/2}\hat{I}_j, \,\,
	\hat{p}_j=C_j^{-1/2}\hat{Q}_j, \,\,
	[\hat{x}_j,\hat{p}_k]={\rm i}\hbar\delta_{jk},
\end{equation}
where $\delta_{jk}$ stands for the Kronecker symbol.

Using (\ref{eq:CanonicalV}), we can rewrite terms of Hamiltonian (\ref{eq:Hamilt0}) in the form:
\begin{equation}
	\hat{\mathcal{H}}_j=\frac{\hat{p_j}^2}{2}+\frac{\omega_j^2\hat{x_j}^2}{2}, \quad \hat{V}=g\omega_1\omega_2\hat{x}_1\hat{x}_2, 
\end{equation}
where $g=L_{12}(L_1L_2)^{-1/2}$ is the dimensionless coupling constant. 

One can diagonalize Hamiltonian (\ref{eq:Hamilt0}) using the canonical transformation to new canonical variables \cite{HamiltDiag}:
\begin{eqnarray}\label{eq:NewCoord}
\begin{split}
	\begin{pmatrix}
	\hat{X}_1 \\
	\hat{X}_2 \\
	\end{pmatrix}=
	M(\Theta)
	\begin{pmatrix}
	\hat{x}_1 \\
	\hat{x}_2 \\
	\end{pmatrix}, \quad
	\begin{pmatrix}
	\hat{P}_1 \\
	\hat{P}_2 \\
	\end{pmatrix}=
	M(\Theta)
	\begin{pmatrix}
	\hat{p}_1 \\
	\hat{p}_2 \\
	\end{pmatrix}, 
\end{split}
\end{eqnarray}
where $M(\Theta)=U(\Theta,0)$ is the rotation operator and
$$
	\Theta=\arctan{\frac{2g\omega_1\omega_2}{\omega_1^2-\omega_2^2}}.
$$

As a result, we obtain Hamiltonian (\ref{eq:Hamilt0}) in the form of two unit mass non-interacting oscillators,
\begin{equation} \label{eq:Hamilt1}
	\hat{\mathcal{H}}=\frac{\hat{P_1}^2}{2}+\frac{\Omega_1^2\hat{X_1}^2}{2}+\frac{\hat{P_2}^2}{2}+\frac{\Omega_2^2\hat{X_2}^2}{2},
\end{equation}
with new resonant frequencies,
$$
\begin{aligned}
	\Omega_1^2=\omega_1^2\cos^2\Theta+\omega_2^2\sin^2\Theta+g\omega_1\omega_2\sin2\Theta, \\
	\Omega_2^2=\omega_1^2\sin^2\Theta+\omega_2^2\cos^2\Theta-g\omega_1\omega_2\sin2\Theta.
\end{aligned}
$$

As a computational basis for our further consideration, we choose the eigenstates $\ket{m,n}$ of the Hamiltonian $\hat{\mathcal{H}}_1+\hat{\mathcal{H}}_2$, such that,
$$
	\left(\hat{\mathcal{H}}_1+\hat{\mathcal{H}}_2 \right) \ket{m,n} = E_{m,n} \ket{m,n}, \quad m,n=0,1,2,\ldots
$$
with $E_{m,n}=\hbar\omega_1\left(m+{1}/{2} \right)+\hbar\omega_2\left(n+{1}/{2} \right)$, where $m$ and $n$ are numbers of energy quanta of the oscillators.

The eigenstates of Hamiltonian (\ref{eq:Hamilt1}) we denote with tildes over integer numbers ({\it e.g}, $\ket{\tilde{1},\tilde{2}}$),
$$
	\left(\hat{\mathcal{H}}_1+\hat{\mathcal{H}}_2+\hat{V}\right)
	\ket{\tilde{m},\tilde{n}}
	=E_{\tilde{m},\tilde{n}} \ket{\tilde{m},\tilde{n}}, \quad m,n=0,1,2,\ldots
$$
with 
$E_{\tilde{m},\tilde{n}}=\hbar\Omega_1\left(m+{1}/{2} \right) + \hbar\Omega_2\left(n+{1}/{2}\right)$.

The coefficients of decomposition of these states in the computational basis we denote as follows:
$$
	C_{i,j}^{m,n}=\bra{i,j}\tilde{m},\tilde{n}\rangle.
$$ 
One can easily calculate them using the wave function of the harmonic oscillator eigenstates,
\begin{multline} \label{eq:Harm}
	\Psi_{l}^{(\Omega)}(X){=}\frac{\Omega^{1/4}}{\sqrt{(2^ll!)\sqrt{\pi}}}\exp{\left( -\frac{1}{2}\Omega X^2\right)}\mathbf{H}_l\left(\sqrt{\Omega}X\right),
\end{multline}
where $m$ is the corresponding quantum number, $X$ is the coordinate, $\Omega$ is the frequency, and $\mathbf{H}_l$ stands for the Hermitian polynomial of $l$-th order. 
Thus, we obtain:
\begin{multline} \label{eq:Decomp}
	C_{i,j}^{m,n}=\iint_{-\infty}^\infty
	\mathrm{d}x_1\mathrm{d}x_2
	\Psi_{i}^{(\omega_1)}(x_1) \ 
	\Psi_{j}^{(\omega_2)}(x_2) \times \\ \times
	\Psi_{m}^{(\Omega_1)}(X_1) \ 
	\Psi_{n}^{(\Omega_2)}(X_2),
\end{multline}
Here, the complex conjugation is omitted because there is no imaginary part in the considered wave functions. 
We note that function (\ref{eq:Harm}) is even function for even $m$, and it is odd function otherwise; thus, the parity implies that $C_{i,j}^{m,n}=0$ for odd $i+j+m+n$.

\begin{figure*}[t]
\begin{centering}
\includegraphics[width=0.9\textwidth]{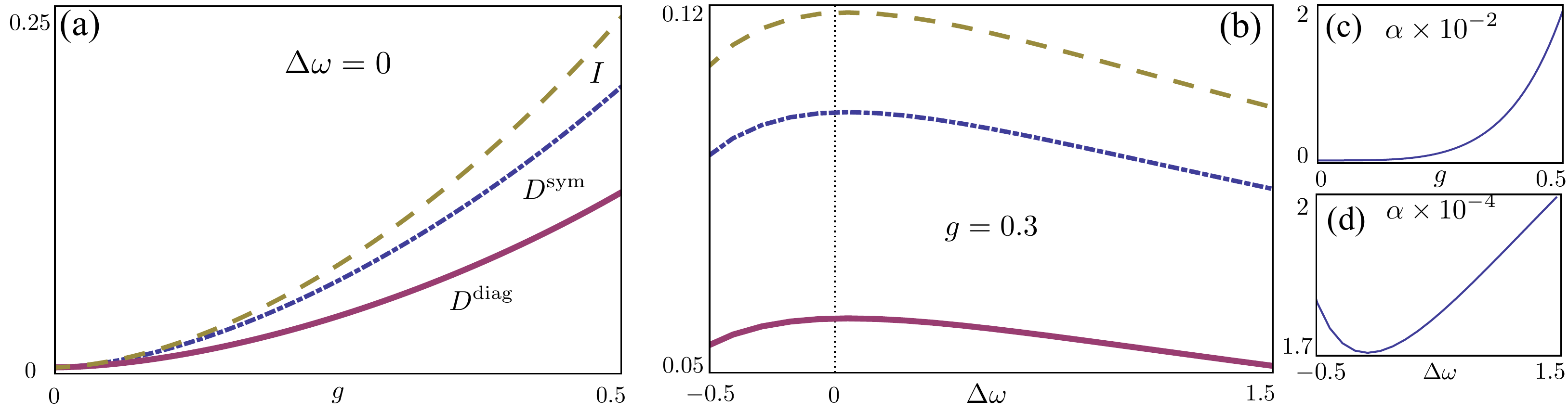}
\end{centering}
\vskip -4mm
\caption{
Measures of correlations in two-qubit approximation of ground state (\ref{eq:Gr2qb}):
quantum mutual information $I$ (dashed), 
symmetric discord $D^\sym$ (dot-dashed), and
diagonalizing discord $D^\diag$ (solid).
In (a) the measures of correlations as functions of the coupling parameter $g$ at the detuning $\domega=0$;
in (b) the same measures as functions of the detuning $\domega$ at the fixed coupling constant $g=0.3$. 
The validity of the two-qubit approximation estimated by parameter the $\alpha$ (\ref{eq:GrParam}) is shown in (c) and (d), respectively.
For considered state (\ref{eq:Gr2qb}), diagonalizing discord $D^\diag$ is equal to entanglement of formation $\mathcal{E}$ and optimal discord $D^\opt$ (see Eq. (\ref{pure})).} 
\label{fig:GrSt}
\end{figure*}

Further, we consider projections of various states on two-qubit subspace,
$$
	\mathcal{H}^\mathrm{2qb}=\mathrm{span}\{\ket{0,0},\ \ket{0,1},\ \ket{1,0},\ \ket{1,1}\},
$$
with the projector operator 
$$
	\Ptqb:\mathfrak{S}(\mathcal{H})\to\mathfrak{S}(\mathcal{H}^\tqb),
$$
in the form:
$$
	\Ptqb=\ketbraa{0}{0}+\ketbraa{0}{1}+\ketbraa{1}{0}+\ketbraa{1}{1}.
$$
Thus, for each state $\hat{\rho}$ we can obtain its two-qubit approximation as follows:
$$
	\hat{\rho}^\tqb=\frac{\Ptqb \hat{\rho} \Ptqb}{\mathrm{Tr}[\Ptqb \hat{\rho} \Ptqb]}.
$$

The accuracy of the approximation can be estimated by the parameter,
\begin{equation} \label{eq:alpha}
	\alpha=1-\mathrm{Tr}[\Ptqb \hat{\rho} \Ptqb]\geq0,
\end{equation}
which is, obviously, zero for states $\hat{\rho}\in\mathfrak{S}(\mathcal{H}^\tqb)$.

\subsection{The ground state}

The ground state of Hamiltonian (\ref{eq:Hamilt1}) has the simple form $\ket{\psi_\mathrm{gr}}=\ket{\tilde{0},\tilde{0}}$.
Due to the parity, its projection on the subspace $\mathcal{H}^\mathrm{2qb}$ consists only of two terms:
\begin{equation}\label{eq:Comp}
	\Ptqb\ket{\tilde{0},\tilde{0}}= C^{00}_{00}\ket{0,0}+C^{00}_{11}\ket{1,1}.
\end{equation}
Then, the two-qubit approximation of the state $\hat{\rho}_\mathrm{gr}=\ket{\psi_\mathrm{gr}}\bra{\psi_\mathrm{gr}} $ reads:
\begin{equation} \label{eq:Gr2qb}
\hat{\rho}_\mathrm{gr}^\tqb=\frac{1}{\left(C^{00}_{00}\right)^2+\left(C^{00}_{11}\right)^2}
\begin{pmatrix}
	(C^{00}_{00})^2 & 0 & 0 & C^{00}_{00}C^{00}_{11} \\
	0 & 0 & 0 & 0 \\
	0 & 0 & 0 & 0 \\
	C^{00}_{00}C^{00}_{11} & 0 & 0 & (C^{00}_{11})^2
\end{pmatrix}
\end{equation}
with accuracy parameter (\ref{eq:alpha}) in the form:
\begin{equation} \label{eq:GrParam}
	\alpha=1-\left(C^{00}_{00}\right)^2-\left(C^{00}_{11}\right)^2.
\end{equation}
As all pure states, \eqref{eq:Gr2qb} belongs to the class of $\X$-states.

\subsection{The thermal state}

The second state we are interested in is a thermal state. 
It is given by the general expression,
$$
	\hat{\rho}_\mathrm{th}=\frac{1}{Z}\exp{\left( -\frac{\hat{\mathcal{H}}}{\kb T}\right)}, \quad Z=\mathrm{Tr}\left[\exp{\left(-\frac{\hat{\mathcal{H}}}{\kb T}\right)}\right]
$$
where $\hat{\mathcal{H}}$ is Hamiltonian (\ref{eq:Hamilt1}), and $Z$ is the partition function.
More explicitly, the density operator can be written in the following form:
\begin{equation}\label{eq:ThSt}
	\hat{\rho}_\mathrm{th}=\frac{1}{Z}\sum_{m,n\geq 0}\left[\exp{\left( -\frac{E_{\tilde{m},\tilde{n}}}{\kb T}\right) }\ketbra{m}{n}\right].
\end{equation}

We assume that the temperature $T$ is low enough to consider final number of terms in (\ref{eq:ThSt}) in a such way that the total number of quanta in each term is non greater than two. 
Again, we use projections on the subspace $\mathcal{H}^\mathrm{2qb}$.
The final form of considered state (\ref{eq:ThSt}) reads:
\begin{equation}\label{eq:ThSt2}
	\hat{\rho}_\mathrm{th}^\mathrm{2qb}=\frac{1}{Z_1}\hat{W}, \qquad
	Z_1=\mathrm{Tr}\left[\hat{W}\right],
\end{equation}
where the operator $\hat{W}$ has the form:
$$
	\hat{W}=\Ptqb\sum_{\substack{m,n\geq 0, \\m+n\leq2}}\left[\exp{\left( -\frac{E_{\tilde{m},\tilde{n}}}{\kb T}\right) }\ketbra{m}{n}\right]\Ptqb.
$$

In this case, accuracy parameter (\ref{eq:alpha}) is given by:
\begin{equation}\label{eq:ThParam}
	\alpha=1-\frac{Z_1}{Z}.
\end{equation}
From the parity, it follows that nonzero projections of vectors used in~\eqref{eq:ThSt} are given by [see also \eqref{eq:Comp}]:
$$
\begin{aligned}
	\hat{\Pi}^\mathrm{2qb}\ket{\tilde{0},\tilde{1}}&=C^{01}_{01}\ket{0,1}+C^{01}_{10}\ket{1,0}, \\
	\hat{\Pi}^\mathrm{2qb}\ket{\tilde{1},\tilde{0}}&=C^{10}_{01}\ket{0,1}+C^{10}_{10}\ket{1,0}, \\
	\hat{\Pi}^\mathrm{2qb}\ket{\tilde{1},\tilde{1}}&=C^{11}_{00}\ket{0,0}+C^{11}_{11}\ket{1,1}, \\
	\hat{\Pi}^\mathrm{2qb}\ket{\tilde{0},\tilde{2}}&=C^{02}_{00}\ket{0,0}+C^{02}_{11}\ket{1,1}, \\
	\hat{\Pi}^\mathrm{2qb}\ket{\tilde{2},\tilde{0}}&=C^{20}_{00}\ket{0,0}+C^{20}_{11}\ket{1,1}. \\
\end{aligned}
$$
This fact implies that state $\hat{\rho}_\mathrm{th}^\mathrm{2qb}$ (\ref{eq:ThSt}) is the $\X$-state.

\begin{figure*}[t]
\begin{centering}
\includegraphics[width=1\textwidth]{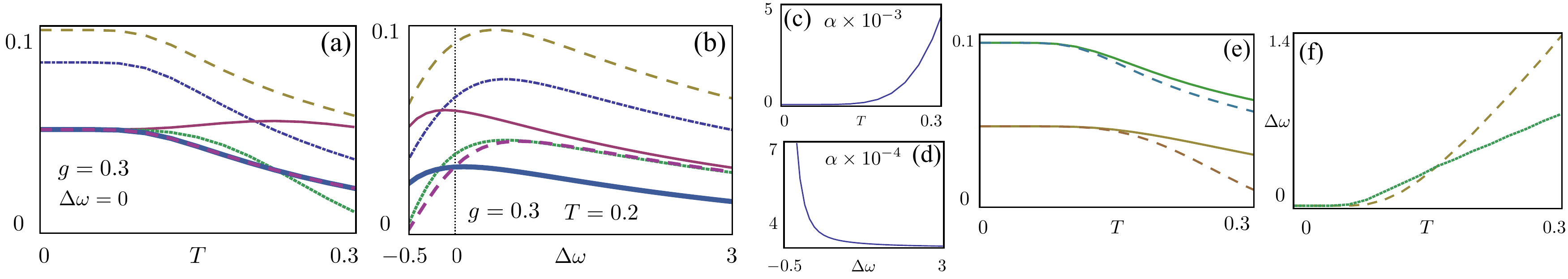}
\end{centering}
\vskip -4mm
\caption
{
Measures of correlations in two-qubit approximation of thermal state (\ref{eq:ThSt2}):
quantum mutual information $I$ (dashed), 
symmetric discord $D^\sym$ (dot-dashed), 
diagonalizing discord $D^\diag$ (solid),
canonical discord $D^{(A)}$ (bold solid),
canonical discord $D^{(B)}$ (bold dashed), and
entanglement of formation $\mathcal{E}$ (dotted).
In (a), the measures of correlations as functions of the temperature $T$ at the resonance $\domega=0$ and the coupling constant $g=0.3$.
In (b) the measures of correlations as functions of the detuning $\domega$ at the coupling constant $g=0.3$ and the temperature $T=0.2$. 
The validity of two-qubit approximation estimated by parameter $\alpha$ given by (\ref{eq:ThParam}) is shown in (c) and (d), respectively.
In (e) the maximum possible values of $I$ and $\mathcal{E}$ accessible at given temperature and the coupling constant $g=0.3$ (solid lines), 
compared with corresponding values (dashed lines) at the resonance $\domega=0$. 
In (f) values of the detuning $\domega$, which correspond to the maximum values of the quantum mutual information $I$ (dashed) and the entanglement of formation $\mathcal{E}$ (dotted).
} 
\label{fig:ThSt}
\end{figure*}

\subsection{Quantum discord for two coupled circuits}

In this part, we consider different measures of correlations for $\X$-states, realized in two-qubit approximations of ground (\ref{eq:Gr2qb}) and thermal states (\ref{eq:ThSt2}) of two coupled oscillators. 
The considered measures of correlations are: 

(i) quantum mutual information (\ref{eq:QMutInf}); 

(ii) diagonalizing discord $D^\diag$ (\ref{eq:DiagTomDisc}); 

(iii) symmetrizing discord $D^\sym$ (\ref{eq:SymTomDisc});

(iv) canonical quantum discords $D^{(A)}$ and $D^{(B)}$ obtained in form (\ref{eq:Disc}) with measurements on the subsystems $A$ and $B$, correspondingly;

(iv) entanglement of formation $\mathcal{E}$. 

The latter we compute using the concurrence $\mathcal{C}$ via general formula \cite{concurrence}:
$$
	\mathcal{E}=h\left( \frac{1}{2}+\frac{1}{2}\sqrt{ 1-\mathcal{C}^2 }\right),
$$
where $h(x)=-x\log x-(1-x)\log(1-x)$ is the binary entropy function. 
In turn, the concurrence $\mathcal{C}$ for the two-qubit density matrix $\rho_{AB}$ is given by:
$$
	\mathcal{C}=\max{\left( 0,\ \lambda_1-\lambda_2-\lambda_3-\lambda_4\right) },
$$
where $\lambda_i$-s are written in the descending order eigenvalues of the following matrix:
$$
	R=
	\sqrt{\sqrt{\rho}\left( \sigma_y\otimes\sigma_y
	\right)
	\rho^{*}
	\left( 
	\sigma_y\otimes\sigma_y
	\right) \sqrt{\rho}}.
$$
Here, a star stands for complex conjugation, $\sigma_y$ for the corresponding Pauli matrix.

We use dimensionless variables, assuming $\hbar=\kb=1$.
Without loss the generality, the frequency $\omega_1$ of the first circuit is assumed to be $1$, and we express the second one via the detuning $\domega$: $\omega_2=\omega_1+\domega$.  
We consider two types of detuning: the blue detuning ($\domega>0$) and the red detuning ($\domega<0$).

We note that optimal discord $D^\opt$ is the minimum of values between $D^\diag$ and $D^\sym$ (see (\ref{eq:OptD})).
In addition, to reveal an extent in which our two-qubit approximation keeps to be appropriate, we compute accuracy parameters \eqref{eq:alpha} and \eqref{eq:ThParam}.

\subsubsection{The ground state}

We start with investigation of the two-qubit approximation of ground state (\ref{eq:Gr2qb}). 
First, for the resonance case ($\domega=0$) and different values of the coupling constant $g$, obtained results are presented in Fig. 3(a).
We see that all measures of correlations grow with the coupling constant $g$. 
As for all pure states (see (\ref{pure})), tomographic discord calculated in diagonalizing scheme $D^\diag$ appears to be optimal ($D^\opt=D^\diag$), and
it coincides with canonical discords $D^{(A)}$ and $D^{(B)}$ as well as with the entanglement of formation $\mathcal{E}$. 
As state (\ref{eq:Gr2qb}) is not maximally entangled, symmetric discord  $D^\sym$ is invariably greater than diagonal discord $D^\diag$. 
In other words, equality (\ref{pure}) holds.

Second, Fig. 3(b) shows an influence of detuning at fixed value of the coupling constant ($g{=}0.3$). 
Level of correlations decreases but in rather small extent. 
As for all non maximally entangled states, the equality $D^\opt=D^\diag$ holds. 

The validity of the two-qubit approximation estimated by parameter (\ref{eq:GrParam}) is presented on Fig. 3(c)--(d), respectively.
We note that the variation of accuracy parameter $\alpha$ (\ref{eq:GrParam}) for detuning $\Delta\omega\ne0$ is sufficiently small. 

\subsubsection{The thermal state}

Here, we study the measures of correlations in two-qubit approximation for thermal state (\ref{eq:ThSt}). 

First, for the resonance case ($\domega=0$) and the coupling constant $g=0.3$, the results for the measures of correlations as functions of the temperature are presented in Fig. 4(a).
We note that in the resonant case, the both canonical discords (\ref{eq:Disc}) are equal ($D^{(A)}=D^{(B)}$). 
Furthermore, at the temperature $T\lesssim 0.1$, they are very close to two other correlation measures $D^\opt=D^\diag$ and $\mathcal{E}$.
At the temperature $T\gtrsim0.1$, diagonalizing discord $D^\diag$ becomes an inappropriate measure of quantum correlations; and at $T\gtrsim0.22$, optimal measurement scheme turns from the diagonalizing to the symmetrizing one. 
At higher temperatures, the value of $D^\opt=D^\sym$ naturally decays as all other measures. 

Second, at the fixed temperature $T=0.3$ and the coupling constant $g=0.3$, the influence of the detuning $\domega$ depicted in Fig. 4(b).
We see that with the presence of detuning equality between canonical discords fails.
Thus, we conclude:
\begin{eqnarray}
	D^{(A)}>D^{(B)}, \quad \mbox{ for}  \quad \domega<0; \nonumber \\
	D^{(B)}>D^{(A)}, \quad \mbox{ for}  \quad \domega>0. \nonumber
\end{eqnarray}
Also, one can mention that maxima of correlations measured by $I$ and $\mathcal{E}$ have shifted to the area with $\domega>0$.
Notably, these maxima correspond to slightly different values of detuning.
The validity of two-qubit approximation estimated by parameter $\alpha$ given by (\ref{eq:ThParam}) is shown in (c) and (d), respectively. 

In Fig. 4(e), at given temperature $T=0.2$ and the coupling constant $g=0.3$, we show the maximum possible values of $I$ and $\mathcal{E}$ and how they compare to corresponding values obtained at resonance.
One can see that the difference for $\mathcal{E}$ is rather dramatic. 
In Fig. 4(f), we show the behavior of the detuning, which give corresponding maximum values of correlation measures.
At high temperature, they have different asymptotic behavior. 
However, the validity of our two-qubit approximation falls, therefore this dependence is not of interest.
As it is expected, the two-qubit approximation validity measured by $\alpha$ decreases with growth of temperature.

Finally, we consider entropic asymmetry of thermal state (\ref{eq:ThSt}) at finite temperature as function of the detuning.
For the temperature $T=0.2$ and the coupling parameter $g=0.3$, obtained results are shown in Fig. 5. 
One can see that entropic asymmetry measured by $d_{AB}$ (\ref{eq:AsMes}) and $d_{AB}^\tom$ (\ref{eq:TomAsMes}) changes its direction {\it exactly} at the resonance.
Furthermore, their behavior is similar to asymmetry between discords $D^{(A)}$ and $D^{(B)}$.
The oscillator with higher frequency (that is $B$ at $\domega>0$; and  $A$ at $\domega<0$) always turns to be an ``effect'' with respect to another oscillator. 
In a mechanical analogy, lower frequency corresponds to higher mass (if we assuming that stiffness coefficients are equal). 
Thus, one can conclude that in thermal states the heavier oscillator, obviously, plays more important role than the lighter one.

Moreover, from comparison with the corresponding behavior of discords $D^{(A)}$ and $D^{(B)}$, 
we conclude that the measurement, made on ``cause'', gives us more access to correlations rather than measurement on ``effect''. 
Speaking in the framework of quantum causal analysis, we have $D^{(\text{``cause''})}<D^{(\text{``effect''})}$.
At the same time, with the growth of asymmetry measured by $d_{AB}$, discord obtained by measurement of the ``effect'' tends to the optimal one.

It is interesting that at high blue detuning ($\domega>0$) $d_{AB}$ becomes larger than unity, that is in principle possible only for quantum systems 
(for classical systems, the highest asymmetry $|d_{AB}|=1$ is obtained when one of the independence function is equal to zero, while the other tends to unity). 
On the other hand, tomographic measure $d_{AB}^\tom$ does not demonstrate such high level of the asymmetry in this state.

\begin{figure}[t]
\begin{center}
\includegraphics[width=0.9\columnwidth]{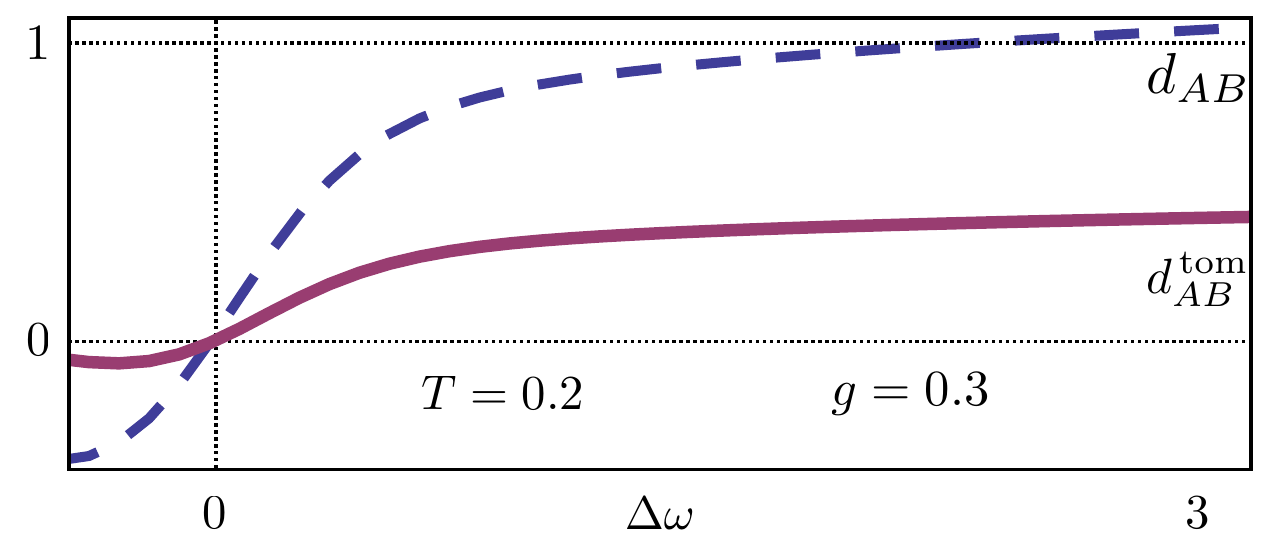}
\end{center}
\vskip -4mm
\caption{Quantum (\ref{eq:AsMes}) and tomographic (\ref{eq:TomAsMes}) measures of entropic asymmetry for thermal state~\eqref{eq:ThSt}.} 
\label{Asym}
\end{figure}

\section{Conclusion and outlook}\label{Conclusion}

In conclusion, we point out the main results of the present paper. 
We have considered tomographic approach to quantum discord and marked out three particular measurement schemes: optimal (\ref{eq:OptTomDisc}), diagonalizing (\ref{eq:DiagTomDisc}), and symmetrizing (\ref{eq:SymTomDisc}) one.
We have established the explicit relation between the tomographic discord, the symmetric discord based on von Neumann measurements, and the measurement-induced disturbance \cite{discord4}.
More precisely, in the general case, the optimal discord coincides with the symmetric discord, while the diagonalizing discord is equal to the measurement-induced disturbance.

We have focused our attention on $\X$-states and have obtained that their optimal discord comes either to diagonalizing or to symmetrizing discord, that implies analytical formula (\ref{eq:OptD}) for its calculation.
Combining this results with quantum causal analysis allows to separate $\X$-states on ``tomographically asymmetric'' and ``tomographically symmetric'' subclasses [Fig. 2(b)].
Numerical results with randomly generated arbitrary mixed states have shown that this separation is the feature of $\X$-states.  

We have considered two-qubit $\X$-states that appear as approximations of ground (\ref{eq:Gr2qb}) and low-temperature thermal states (\ref{eq:ThSt}) of two coupled nanoelectric $LC$-circuits.
We have discussed the robustness of the correlations properties with respect to the environment parameters and have shown that for thermal state (\ref{eq:ThSt}), 
blue detuning of the second circuit can rise the amount of correlations compared to the resonance case.

Also, we have obtained that appeared in coupled circuits $\X$-states mostly belong to ``tomographically asymmetric'' subclass.
Although, variation of the parameters (temperature or detuning) can transform the subclass of $\X$-state from asymmetrical to symmetrical. 
This change is always in the direction where two-qubit approximation becomes inappropriate.

We have shown that the behavior of the entropic asymmetry, as observed by quantum causal analysis, conforms with the behavior of the asymmetry between the canonical discords obtained by measurement on different subsystems. 
In this way, it has been found that a measurement performed on a qubit classified by causal analysis as a ``cause'' gives access to more correlations than a measurement performed on the ``effect''.

We expect the presentation of the results in the framework of a real physical system of two coupled nanoelectric circuits to open the way to experimental study of tomographic quantum discord phenomena.

\section*{Acknowledgements}
We thank S.N. Filippov and Y.V. Kurochkin for useful discussions as well as Y. Huang for insightful comments. 
This work is supported by the Dynasty Foundation, the Council for Grants of the President of the Russian Federation (grant SP-961.2013.5, E.O.K.), and the Russian Foundation for Basic Research (grant 14-08-00606).

\vspace{\baselineskip}
*Corresponding author: {\href{mailto:akf@rqc.ru}{akf@rqc.ru}}

\section*{Appendix I. Tomography of $\X$-states}

To obtain the explicit form of the tomogram for $\X$-state density matrix (\ref{eq:XState}), one needs take the diagonal elements of the matrix:
$$
	\widetilde{\rho}= U(\phi_A,\theta_A)\otimes U(\phi_B,\theta_B) \rho_{AB}^X U^\dagger(\phi_A,\theta_A)\otimes U^\dagger(\phi_B,\theta_B).
$$

In this way, we have the following representation for the tomogram:
\begin{widetext}
$$
\mathcal{T}_{AB}=\left\{
\begin{aligned}
\frac{1}{4}+\frac{1}{4}z_A\cos{\theta_A}+\frac{1}{4}z_B\cos{\theta_B}+\frac{1}{4}z_{AB}\cos{\theta_A}\cos{\theta_B}+\frac{1}{2}\sin{\theta_A}\sin{\theta_B}\left(\rho_{14}\cos{(\phi_A+\phi_B)}+\rho_{23}\cos{(\phi_A-\phi_B)}\right) \\
\frac{1}{4}+\frac{1}{4}z_A\cos{\theta_A}-\frac{1}{4}z_B\cos{\theta_B}-\frac{1}{4}z_{AB}\cos{\theta_A}\cos{\theta_B}-\frac{1}{2}\sin{\theta_A}\sin{\theta_B}\left(\rho_{14}\cos{(\phi_A+\phi_B)}+\rho_{23}\cos{(\phi_A-\phi_B)}\right) \\
\frac{1}{4}-\frac{1}{4}z_A\cos{\theta_A}+\frac{1}{4}z_B\cos{\theta_B}-\frac{1}{4}z_{AB}\cos{\theta_A}\cos{\theta_B}-\frac{1}{2}\sin{\theta_A}\sin{\theta_B}\left(\rho_{14}\cos{(\phi_A+\phi_B)}+\rho_{23}\cos{(\phi_A-\phi_B)}\right) \\
\frac{1}{4}-\frac{1}{4}z_A\cos{\theta_A}-\frac{1}{4}z_B\cos{\theta_B}+\frac{1}{4}z_{AB}\cos{\theta_A}\cos{\theta_B}+\frac{1}{2}\sin{\theta_A}\sin{\theta_B}\left(\rho_{14}\cos{(\phi_A+\phi_B)}+\rho_{23}\cos{(\phi_A-\phi_B)}\right)
\end{aligned}
 \right\},
$$
\end{widetext}
where we use notation
$$
	\begin{aligned}
	z_A=\rho_{11}+\rho_{22}-\rho_{33}-\rho_{44},\\
	z_B=\rho_{11}-\rho_{22}+\rho_{33}-\rho_{44},\\
	z_{AB}=\rho_{11}-\rho_{22}-\rho_{33}+\rho_{44}.
	\end{aligned}
$$

The order of elements in the tomogram is governed by the rules of standard tensor multiplication: 
$$
	\mathcal{T}_{AB}=\left\{{\mathcal{T}_{AB}}_{00},{\mathcal{T}_{AB}}_{01},{\mathcal{T}_{AB}}_{10},{\mathcal{T}_{AB}}_{11} \right\}.
$$
The reduced tomograms takes the simple form:
$$
\begin{aligned}
	\mathcal{T}_{A}&=\left\{\frac{1}{2}+\frac{1}{2}z_A\cos{\theta_A},\frac{1}{2}-\frac{1}{2}z_A\cos{\theta_A}\right\}, \\
	\mathcal{T}_{B}&=\left\{\frac{1}{2}+\frac{1}{2}z_B\cos{\theta_B},\frac{1}{2}-\frac{1}{2}z_B\cos{\theta_B}\right\}
\end{aligned},
$$
without the dependences on $\phi_A$ and $\phi_B$.

\section*{Appendix II. Generation of random two-qubit states}
\subsection{Generation of random $\X$-states}
To generate $\X$-state density matrix (\ref{eq:XState}), we use the following algorithm.
First, we generate the diagonal elements as follows:
$$
	\rho_{ii}=\frac{p_i}{\sum_{j=1}^4p_j}, \qquad p_i=\mathcal{U}(0,1),
$$
where $\mathcal{U}(a,b)$ stands for a uniform distribution on $[a,b]$.

Then we generate off-diagonal elements as:
$$	
	\rho_{14}=\epsilon_{1}\sqrt{\rho_{11}\rho_{44}}, \quad  \rho_{23}=\epsilon_{2}\sqrt{\rho_{22}\rho_{33}}, \quad \epsilon_{1},\epsilon_{2}=\mathcal{U}(0,1).
$$
We note that this rather straightforward algorithm does not generate states uniformly in respect to the Haar measure, 
so it is quite useful for observing two possible subclasses of $X$-states but is not appropriate for a study of a relative volume of each subclass in the whole set of all $X$-states.

\subsection{Generation of random arbitrary two-qubit mixed states}

We generate arbitrary two qubit states in the following form:
$$
	\rho_{AB}=\frac{1}{\sum_{j=1}^{4}p_j} \sum_{k=1}^4 p_k \langle\psi_k|\psi_k \rangle^{-1} |\psi_k\rangle\langle\psi_k|,
$$
where
$$
	|\psi_k\rangle =\begin{pmatrix}
	\mathcal{N}(0,1) \\
	\mathcal{N}(0,1) \\
	\mathcal{N}(0,1) \\
	\mathcal{N}(0,1)
	\end{pmatrix}+\mathrm{i}\begin{pmatrix}
	\mathcal{N}(0,1) \\
	\mathcal{N}(0,1) \\
	\mathcal{N}(0,1) \\
	\mathcal{N}(0,1)
	\end{pmatrix},
	\quad
	p_j=\mathcal{U}(0,1),
$$
with $\mathrm{i}^2=-1$ and $\mathcal{N}(\mu,\sigma)$ being the Gaussian distribution with the expectation value $\mu$ and the standard deviation $\sigma$.
According to Ref. \cite{StateGen}, this method gives a uniform distribution of quantum states.

\end{document}